\documentclass[aps,prd,onecolumn,preprintnumbers,groupedaddress,showpacs,nofootinbib,amssymb]{revtex4}
\usepackage{graphicx,color}
\usepackage{amsmath}
\usepackage{amssymb}
\usepackage{amsfonts}
\usepackage{bm}
\usepackage{cancel}

\allowdisplaybreaks[4]

\begin{document}

\newcommand{\be}{\begin{equation}}
\newcommand{\ee}{\end{equation}}
\newcommand{\bea}{\begin{eqnarray}}
\newcommand{\eea}{\end{eqnarray}}
\newcommand{\nn}{\nonumber \\}
\newcommand{\e}{\mathrm{e}}
\newcommand{\tr}{\mathrm{tr}\,}

\tolerance=5000

\title{Regular Multi-Horizon Black Holes in Modified Gravity with Non-Linear Electrodynamics}

\author{
Shin'ichi Nojiri$^{1,2,3}$\footnote{E-mail address: nojiri@phys.nagoya-u.ac.jp}
and
S.~D.~Odintsov$^{4,5}$\footnote{E-mail address: odintsov@ieec.uab.es}}

\affiliation{
$^1$Department of Physics, Nagoya University, Nagoya 464-8602, Japan \\
$^2$Kobayashi-Maskawa Institute for the Origin of Particles and the Universe,
Nagoya University, Nagoya 464-8602, Japan \\
$^3$KEK Theory Center, IPNS, KEK, Tsukuba, Ibaraki 305-0801, Japan \\
$^4$ICREA, Passeig Luis Companys, 23, 08010 Barcelona, Spain \\
$^5$Institute of Space Sciences (IEEC-CSIC) C. Can Magrans s/n,
08193 Barcelona, Spain
}

\begin{abstract}

We investigated the regular multi-horizon black holes in the Einstein gravity,
$F(R)$ gravity and the 5 dimensional Gauss-Bonnet
gravity, all of them coupled with non-linear electrodynamics.
We presented several explicit examples of the actions which admit the solutions
describing regular black hole space-time with multi-horizons.
Thermodynamics of the obtained black hole solutions is studied. The explicit
expressions of the temperature, the entropy, the thermodynamical energy and
the free energy are obtained.
Although the temperature vanishes in the extremal limit where the radii of the
two horizons
coincide with each other as in the standard multi-horizon black hole
like the Reissner-Nordstr\" om
black hole or the Kerr black hole, the larger
temperature corresponds to the larger horizon radius. This is different from
the standard black holes, where the larger temperature corresponds to the smaller
horizon radius.
We also found that the specific heat
becomes positive for the large temperature, which is also different from the
standard black holes, where the specific heat is negative.
It should be also noted that the thermodynamical energy is not identical with
the ADM mass.
Furthermore in case of the Gauss-Bonnet gravity,it is demonstrated that the
entropy can become negative.

\end{abstract}


\maketitle

\section{Introduction \label{Sec1}}

Usually black hole has both of curvature singularity and horizon(s).
The singularity without horizon might be prohibited by the cosmic censorship
\cite{Penrose:1969pc}.
Conversely, before the concept of the cosmic censorship, it has been found the
space-time with horizon
but without the curvature singularity can be realized by using the non-linear
electromagnetism \cite{JMBardeen:1968}.
This fact iniciated the study of the regular black hole (with non-linear
electrodynamics):
\cite{AyonBeato:1998ub,Bronnikov:2000vy,AyonBeato:2000zs,Elizalde:2002yz,
Dymnikova:2004zc,Nicolini:2005vd,Ansoldi:2006vg,Ansoldi:2008jw,Hossenfelder:2009fc,Johannsen:2015pca,Dymnikova:2015hka,Culetu:2015axa,Ma:2015gpa,Kunstatter:2015vxa,Pradhan:2015wnl,Rodrigues:2016fym,Fan:2016hvf,Chinaglia:2017uqd,Kruglov:2017xmb}.
Recently in \cite{Chinaglia:2017uqd}, the construction of the non-singular
black holes by using the non-linear electrodynamics has been completely
generalized.
The interest to gravity with non-linear electrodynamics is 
caused by several reasons. First of all, non-linear electrodynamics models 
with gravity 
are predicted by  low-energy string effective action. Second, such 
theories may appear as the effective action from quantum gravity coupled 
with matter. Third, such theories may be considered as simple enough models 
mimicking much more complicated fundamental proposals for 
non-perturbative quantum gravity 
with matter at the early universe. It is expected that non-linear 
character of the theory may 
be manifested in quite unusual aspects of black hole thermodynamics.
Specifically, we consider multi-horizon black holes where Nariai limit may 
be easily accessed. This is caused by the well-known fact that such black 
holes  are supposed to be the primordial black holes which were created at 
the early universe. It is expected that some of primordial black holes 
which may have the unusual properties may survive at the current universe. 
If so, they may give us the important information about the properties of 
the early universe. For instance, as we discuss later on, unlike to 
standar one-horizon black holes, the multi-horizon black holes may show up 
not only the Hawking evaporation but also anti-evaporation or related 
instability phenomenon.

The purpose of this work is to study the non-singular black holes in modified
gravity with non-linear electrodynamics.
By using the formulation of Ref.~\cite{Chinaglia:2017uqd} we consider the
problem of non-singular black holes in $F(R)$ gravity which may be responsible for
current acceleratingly
expanding universe (see \cite{Capozziello:2002rd,Carroll:2003wy,Nojiri:2003ft}
and for the review, see
\cite{Nojiri:2006ri,Nojiri:2010wj,Nojiri:2017ncd,Capozziello:2011et})
and 5-dimensional Gauss-Bonnet gravity.
We give some examples of the non-singular black hole space-time with
multi-horizons
and investigate their thermodynamical properties. A motivation why we consider
the 5-dimensional gravity is related with the AdS/CFT correspondence
\cite{Maldacena:1997re,Witten:1998qj}, where the gravity theory in the $D$
dimensional anti-de Sitter (AdS) space-time is equivalent to the conformal
field theory (CFT)
in the $D-1$ dimensional flat space-time, which can be identified with the
boundary of the AdS space-time.
We show, however, that our model does not correspond to any 4-dimensional
relativistic
field theory but the effective 3-dimensional theory or non-relativistic theory.
We also show that there are some regions of the parameter space where the
entropy of 5-dimensional Einstein-Gauss-Bonnet gravity
becomes negative.

It is interesting to discuss the relation between the regular black holes 
and the energy conditions
\cite{Balart:2014jia,Neves:2014aba}.
In the formulation of \cite{Chinaglia:2017uqd}, one can construct any type 
of spherically symmetric
regular black hole and one can check the energy conditions explicitly.
In this paper, we check the conditions for some simple cases.

We often consider the Nariai limit where the radii of two horizons coincide 
with each other.
In some cases, such a black hole is stable due to the phenomena called
anti-evaporation.
It may have quite long life-time although with the radius being small.
This is to compare with small black hole which evaporates in a short time. 
Therefore, the  Nariai black hole may be considered as primordial one and 
might
survive even in the present universe.

In the next section, we give a general formulation to construct the regular
black holes,
especially with multi-horizons, based on Ref.~\cite{Chinaglia:2017uqd}.
Some explicit examples of the black hole space-time with multi-horizons in the
Einstein gravity with non-linear electrodynamics are constructed.
In Section \ref{Sec3}, we generalize the formulation of
Ref.~\cite{Chinaglia:2017uqd} for
$F(R)$ gravity and give the explicit examples of the multi-horizon regular
black hole space-time in the $R^2$ gravity.
It is also demonstrated that one cannot construct any spherically symmetric
regular black hole solutions only by $F(R)$ gravity.
The thermodynamics is also investigated in Section \ref{Sec4})
and some expressions for the entropy $\mathcal{S}$ and the energy $E$ are
found.
It is shown that the obtained expression for the energy $E$ is
often different from
the ADM mass \cite{Arnowitt:1959ah}.
In Section \ref{Sec5}, we consider 5-dimensional Gauss-Bonnet gravity with
non-linear electrodynamics. We develop a formulation to construct the models
which generate the general spherically symmetric space-time and give an example
of the regular black hole with two horizons.
We also investigate the thermodynamics and we find that the entropy can be
negative.
In addition to the entropy $\mathcal{S}$, the expression of the free energy
$F$ is found.
It indicates that the five-dimensional gravity model could not correspond to
any 4-dimensional
relativistic field theory but eventually it corresponds to the effective 3
dimensional theory or non-relativistic theory.
Some summary and outlook are given in the last section.

\section{Regular Black Hole with Non-linear electromagnetism \label{Sec2}}

In this section, we consider the regular black hole with
multi-horizons in the Einstein gravity coupled with non-linear
electromagnetic field.

\subsection{General Formulation of Reconstruction}

In this subsection, we extend the formulation of the reconstruction in
\cite{Chinaglia:2017uqd} so
that we could construct solutions describing the black hole without curvature
singularity but with multi-horizons.

We start from the following action
\begin{equation}
\label{NEM1}
S = \int d^4 x \sqrt{-g} \left\{ \frac{R}{2} - \mathcal{L}\left( I \right)
\right\} \, .
\ee
Here we have chosen the gravitational constant to be unity, $\kappa=1$ and
$I\equiv \frac{1}{4} F^{\mu\nu} F_{\mu\nu}$, with the field strength
$F_{\mu\nu}$ of the
electromagnetic field $A_\mu$,
$F_{\mu\nu} \equiv \partial_\mu A_\nu - \partial_\nu A_\mu$.
Then the equaions of the motion have the following form,
\begin{align}
\label{NEM2}
G_\mu^{\ \nu} =& \frac{d \mathcal{L}}{dI} F_{\alpha\mu} F^{\alpha\nu}
+ \mathcal{L} \delta_\mu^{\ \nu} \, , \\
\label{NEM3}
0=& \nabla^\mu \left( F_{\mu\nu} \frac{d \mathcal{L}}{dI} \right) \, .
\end{align}
We now assume a spherically symmetric static background as in the Schwarzschild
space-time as follows,
\begin{equation}
\label{GBiv}
ds^2 = - \e^{2\nu (r)} dt^2 + \e^{-2\nu (r)} dr^2
+ r^2 \sum_{i,j=1}^2 \tilde g_{ij} dx^i dx^j\, ,
\end{equation}
and $A_i=0$ $\left(i=1,2,3\right)$ and $A_0=A_0(r)$.
Then Eq.~(\ref{NEM3}) has the following form,
\begin{equation}
\label{NEM4}
0 = \frac{d}{dr}\left( r^2 F^{0r} \frac{d \mathcal{L}}{dI} \right) \, .
\end{equation}
By using $I = -\frac{1}{2} \left( F_{0r} \right)^2$, Eq.~(\ref{NEM4}) can be
solved as
\begin{equation}
\label{NEM5}
r^2 \frac{d \mathcal{L}}{dI} = \frac{Q}{\sqrt{ - 2I}}\, .
\end{equation}
We now define a new variable $X$ by
\begin{equation}
\label{NEM6}
X \equiv Q \sqrt{ - 2I}\, .
\end{equation}
Then in \cite{Chinaglia:2017uqd}, the following equations have been obtained,
\begin{align}
\label{NEM7}
\frac{d}{dr} \left( r \left( \e^{2\nu} - 1 \right) \right) = - r^2 \rho \, , \\
\label{NEM8}
\mathcal{L} = \frac{X}{r^2} - \rho \, , \\
\label{NEM9}
X = - \frac{r^3}{2} \frac{d\rho}{dr}\, .
\end{align}
Here $\rho$ is the energy-density of the non-linear electromagnetic field
obtained from the Lagrangian density $\mathcal{L}$.
For the geometry expressed by $\e^{2\nu(r)}$ in (\ref{GBiv}), by using (\ref{NEM7}),
we find the explicit $r$-dependence of $\rho$.
Then by using (\ref{NEM9}), one can express $X$ in terms of $r$, which can be
solved with respect to $r$ as $r=r \left( X \right)$.
Substituting the obtained expression $r=r \left( X \right)$ into (\ref{NEM8}),
we obtain the
explicit form of $\mathcal{L}$, $\mathcal{L}=\mathcal{L} \left( X \right)$.
In general, however, one cannot solve Eq.~(\ref{NEM9}) explicitly with respect
to $r$ as a function of
$X$ and also there might not be the one-to-one correspondence between $X$ and
$r$.
In this case, one cannot obtain the Lagrangian density (\ref{NEM8}).
In order to avoid this problem, we may introduce the auxiliary fields $B$ and
$C$.
It is straightforward to obtain the $r$-dependence of the energy-density
$\rho$ by using (\ref{NEM7})
for a given $\e^{2\nu(r)}$, $\rho=\rho(r)$.
Then by using the auxiliary fields $B$ and $C$, we may consider the following
Lagrangian density
$\mathcal{L}_{BCX}$ instead of $\mathcal{L}$,
\begin{equation}
\label{LBCX}
\mathcal{L} \to
\mathcal{L}_{BCX} \equiv \frac{X}{C^2} - \rho \left(r=C\right)
+ B \left( X + \frac{C^3}{2} \left. \frac{d\rho}{dr} \right|_{r=C} \right) \, .
\end{equation}
By the variation over $B$, we obtain (\ref{NEM9}) with $r=C$.
In case that $C$ can be solved with respect to $X$, by substituting the
obtained expression
into (\ref{LBCX}), one gets the Lagrangian density in (\ref{NEM8}).

\subsection{General Regular Black Hole with Multi-Horizons}

We now consider the general regular multi-horizon black hole, where there is no
curvature singularity.
In the background (\ref{GBiv}), the scalar curvature is given by
\begin{equation}
\label{ScalarR}
R = \e^{2\nu}\left[ - 2\nu'' - 4 {\nu'}^2 - \frac{8\nu'}{r}
+ \frac{2 \e^{- 2\nu} - 2}{r^2} \right] \, .
\end{equation}
Therefore in order to avoid the curvature singularity, we find $\e^{2\nu}\to 1$
in the limit of
$r\to 0$.
Furthermore because $\partial_\mu \e^{2\nu}$ is a single-valued function, we
also require
$\e^{2\nu}=1 + \mathcal{O}\left( r^2 \right)$ ar $r\to 0$.
We also require $\nu''$ does not diverge at $r=0$.

If the background is the Minkowski space-time or the anti-de Sitter space-time,
one may express $\e^{2\nu(r)}$ as follows,
\begin{equation}
\label{NEM10}
\e^{2\nu(r)} = \frac{ (r-r_1)(r-r_2) \cdots (r - r_{2N})}{h(r)}\, .
\end{equation}
Here $N$ is a positive integer, $h(r)$ does not vanish
and $h(r)\to r_1 r_2 \cdots r_{2N} \left( 1 - \left( \sum_{i=1}^{2N}
\frac{1}{r_i} \right) r
+ \mathcal{O}\left( r^2 \right) \right)$ in the limit of $r\to 0$.
Therefore $\e^{2\nu(r)}$ behaves as $\e^{2\nu(r)}\to 1
+ \mathcal{O}\left(r^2\right)$
and therefore the space-time is regular at $r=0$.
On the other hand, in the limit of $r\to \infty$, $h(r)$ behaves as $h(r)\to
r^{2N}$ for the
Minkowski background or $h(r)\to \frac{r^{2N+2}}{l^2}$ with a length parameter
$l$ for the
anti-de Sitter background.
In (\ref{NEM10}, there are $2N$ horizons at $r=r_1$, $r_2$, $\cdots$, and
$r_{2N}$.
We may assume $r_1 \leq r_2 \leq \cdots \leq r_{2N}$.
On the other hand, in the de Sitter background, $h(r)$ could be given by
\begin{equation}
\label{NEM11}
\e^{2\nu(r)} = \frac{ (r-r_1)(r-r_2) \cdots (r - r_{2N-1})}{h(r)}\, .
\end{equation}
Here, again, $N$ is a positive integer, $h(r)$ does not vanish.
In the limit of $r\to 0$, we require $h(r)\to - r_1 r_2 \cdots r_{2N-1}
\left( 1 - \left( \sum_{i=1}^{2N-1} \frac{1}{r_i} \right) r
+ \mathcal{O}\left( r^2 \right) \right)$
and therefore $\e{2\nu(r)}$ behaves as $\e^{2\nu(r)}\to 1 +
\mathcal{O}\left(r^2\right)$, again
and the space-time is regular at $r=0$.
On the other hand, in the limit of $r \to \infty$, we also require $h(r)$
behaves as
$h(r)\to - \frac{r^{2N+1}}{l^2}$ with a length parameter $l$.
In (\ref{NEM11}), there are $2N-1$ horizons at $r=r_1$, $r_2$, $\cdots$, and
$r_{2N-1}$
and if we assume $r_1 \leq r_2 \leq \cdots \leq r_{2N-1}$, $r=r_{2N-1}$
corresponds to the
cosmological horizon.
This may show that the regular black hole in the anti-de Sitter space-time or
the Minkowski space-time
has even-number of horizons.
On the other hand, the regular black hole in de Sitter space-time has odd
number of horizons,
in which the largest horizon corresponds to the cosmological horizon.

\subsection{Example in the Minkowski Background}

We consider the model
\begin{align}
\label{NNem1}
& \e^{2\nu} = 1 - \frac{\alpha r^2}{\beta + r^3}
= \frac{ \left( r + r_0 \right) \left( r - r_1 \right) \left( r - r_2\right) }{\beta + r^3}  \, , \nn
& \alpha = \frac{r_1^2 + r_2^2 + r_1 r_2}{r_1 + r_2} \, , \quad
\beta = \frac{r_1^2 r_2^2}{r_1 + r_2} \, , \quad r_0 = \frac{r_1 r_2}{r_1 + r_2} \, .
\end{align}
Then we find
\begin{equation}
\label{NNem2}
\rho = \frac{3 \alpha \beta}{\left( \beta + r^3 \right)^2}>0 \, , \quad
X = \frac{ 9\alpha \beta r^5}{\left( \beta + r^3 \right)^3} \, , \quad
\mathcal{L} = \frac{3\alpha \beta \left( 2r(X)^3 - \beta \right)}{\left( \beta
+ r(X)^3 \right)^3} \, .
\end{equation}
By using the auxiliarry field $B$ and $C$, one gets
\begin{equation}
\label{NNem3}
\mathcal{L}_{BCX} = \frac{3\alpha \beta \left( 2C^3 - \beta \right)}{\left( \beta + C^3 \right)^3}
+ B \left( X -  \frac{ 9\alpha \beta C^5}{\left( \beta + C^3 \right)^3} \right) \, .
\end{equation}
Then redefining $C\to \beta^{\frac{1}{3}} C$, $A_\mu \to Q^{-1} \beta^{\frac{8}{3}} A_\mu$
$\left( X \to \beta^{\frac{8}{3}} X \right)$, $B\to \beta^{-\frac{8}{3}}$, and $\alpha \to \beta \tilde \alpha$, we can rewrite the Lagrangian density  (\ref{NNem3}) as follows,
\begin{equation}
\label{NNem4}
\mathcal{L}_{BCX} = \frac{3\tilde\alpha \left( 2C^3 - 1 \right)}{\left( 1 + C^3 \right)^3}
+ B \left( X -  \frac{ 9\tilde\alpha C^5}{\left( 1 + C^3 \right)^3} \right) \, .
\end{equation}
Therefore the only parameter in the theory is
\begin{equation}
\label{NNem5}
\tilde\alpha = \frac{\alpha}{\beta} = \frac{r_1^2 + r_2^2 + r_1 r_2}{r_1^2 r_2^2} \, .
\end{equation}
Then we can choose the extremal limit where
\begin{equation}
\label{NNem6}
r_1, r_2 \to \sqrt{\frac{3}{\tilde\alpha}} \, .
\end{equation}
One may consider the energy conditions,
\begin{eqnarray*}
&\circ &\ \mbox{Null Energy Condition (NEC):} \rho + p_r \geq 0 \ \mbox{and}\
\rho + p_T \geq 0 \\
&\circ &\ \mbox{Weak Energy Condition (WEC):}
\rho\geq 0 \, , \ \rho + p_r \geq 0 \, , \quad
\rho + p_T \geq 0\\
\label{phtm9}
&\circ &\ \mbox{Strong Energy Condition (SEC):}
\rho + p_r + 2 p_T \geq 0\, , \ \rho + p_r \geq 0 \, , \  \mbox{and}\
\rho + p_T \geq 0 \\
\label{phtm10}
&\circ &\ \mbox{Dominant Energy Condition (DEC):}
\rho\geq 0 \, , \ \rho \pm p_r \geq 0 \, , \  \mbox{and}\
\rho \pm p_T \geq 0
\end{eqnarray*}
Here $p_r$ is that the pressure for the radial direction and
$p_T$ is the pressure for the angle direction.
\begin{align}
\label{NNem2B}
& \rho = \frac{X}{r^2} - \mathcal{L}
= \frac{3 \alpha \beta}{\left( \beta + r^3 \right)^2}>0 \, , \quad
p_r = \frac{X}{r^2} + \mathcal{L} \, , \quad
p_T = \mathcal{L} = \frac{3\alpha \beta \left( 2r^3 - \beta \right)}{\left( \beta
+ r^3 \right)^3} \, , \quad
X = \frac{ 9\alpha \beta r^5}{\left( \beta + r^3 \right)^3} >0 \, .
\end{align}
Then we find,
\begin{equation}
\label{NNem14}
\rho + p_r = \frac{2X}{r^2}>0\, , \quad \rho + p_T = \frac{X}{r^2} > 0\, .
\end{equation}
Therefore the Null Energy Condition and the Weak Energy Condition are satisified.
On the other hand, because
\begin{equation}
\label{NNem15}
\rho + p_r + 2 p_T = \frac{2X}{r^2} + 2 \mathcal{L}
= \frac{ 6\alpha \beta \left( 5 r^3 - \beta \right)}{\left( \beta + r^3 \right)^3}\, ,
\end{equation}
the Strong Energy Condition is not satisfied in the region
\begin{equation}
\label{NNem16}
r < \left( \frac{\beta}{5} \right)^{\frac{1}{3}}
= \left( \frac{r_1^2 r_2^2}{5 \left( r_1 + r_2 \right)}\right)^{\frac{1}{3}}\, .
\end{equation}
We also find
\begin{equation}
\label{NNem17}
\rho - p_r = - 2 \mathcal{L}
= \frac{6\alpha \beta \left( - 2r^3 + \beta \right)}{\left( \beta + r^3 \right)^3} \, , \quad
\rho - p_T = \frac{3\alpha \beta \left( - r^3 + 2 \beta \right)}
{\left( \beta + r^3 \right)^3} \, .
\end{equation}
Therefore the Dominant Energy Condition is broken in the region
\begin{equation}
\label{NNem18}
r < \left( 2\beta \right)^{\frac{1}{3}}
= \left( \frac{2r_1^2 r_2^2}{r_1 + r_2}\right)^{\frac{1}{3}}\, .
\end{equation}

Thus, we demonstrated that black hole configuration under consideration 
may fulfil the 
energy conditions.

\subsection{Example in the anti-de Sitter Background}

As an example of the regular black hole where there appear two horizons in the
anti-de Sitter background, we consider the following one,
\begin{equation}
\label{NEM12B}
\e^{2\nu(r)} = \frac{(r^2 - r_1^2)(r^2 - r_2^2)}{r_1 r_2 \left( r^2 + r_1r_2
\right)}
= \frac{( r-r_1 )(r - r_2)}{h(r)} \, , \quad
h(r) = \frac{r_1 r_2 \left( r^2 + r_1r_2 \right)}{(r + r_1)(r + r_2)} \, .
\end{equation}
Then the length parameter in the anti-de Sitter space-time is given by $l^2=r_1
r_2$.
By using Eqs.~(\ref{NEM7}) and (\ref{NEM9}), we find
\begin{align}
\label{NEM13}
\rho =& - \frac{ 2 \left\{ 3r^4 + \left( 4 r_1 r_2 - r_1^2 - r_2^2 \right) r^2
  - 3 r_1^2 r_2^2
  - 3 r_1^3 r_2 - 3 r_1 r_2^3\right\} }{r_1 r_2 \left( r^2 + r_1 r_2 \right)^2}
\, , \\
\label{NEM13b}
X =& - \frac{2r^4 \left\{ - 8 r^6 + \left( - 4 r_1 r_2 + r_1^2 + r_2^2 \right)
r^4
+ \left( 6 r_1^2 r_2^2 + 5 r_1^3 r_2 + 5 r_1 r_2^3 \right) r^2
+ 10 r_1^3 r_2^3 + 5 r_1^4 r_2^2 + 5 r_1^2 r_2^4 \right\}}{
r_1 r_2 \left( r^2 + r_1 r_2 \right)^3} \, .
\end{align}
Because we cannot solve Eq.~(\ref{NEM13b}) with respect to $r$,
we use the Lagrangian density in (\ref{LBCX}),
\begin{align}
\label{NEM13c}
& \mathcal{L}_{BCX} = \frac{X}{C^2}
  - \frac{ 2 \left\{ 3C^4 + \left( 4 r_1 r_2 - r_1^2 - r_2^2 \right) C^2 - 3
r_1^2 r_2^2
  - 3 r_1^3 r_2 - 3 r_1 r_2^3\right\} }{r_1 r_2 \left( C^2 + r_1 r_2 \right)^2}
\nn
&+ B \left[ X + \frac{2 C^4 \left\{ - 8 C^6 + \left( - 4 r_1 r_2 + r_1^2 +
r_2^2 \right) C^4
+ \left( 6 r_1^2 r_2^2 + 5 r_1^3 r_2 + 5 r_1 r_2^3 \right) C^2
+ 10 r_1^3 r_2^3 + 5 r_1^4 r_2^2 + 5 r_1^2 r_2^4 \right\}}{
r_1 r_2 \left( C^2 + r_1 r_2 \right)^3} \right] \, .
\end{align}
We now redefine,
\begin{equation}
\label{redefinitions}
C^2 \to r_1 r_2 C^2\, , \quad X \to r_1 r_2 X \, , \quad B \to \frac{B}{r_1
r_2}\, ,
\quad \alpha \equiv \frac{r_1^2 + r_2^2}{r_1 r_2}\, , \quad l^2 = r_1 r_2\, .
\end{equation}
Then the Lagrangian density (\ref{NEM13c}) can be rewritten as
\begin{align}
\label{NEM13cc}
\mathcal{L}_{BCX} =& \frac{X}{C^2}
  - \frac{ 2 \left\{ 3C^4 + \left( 4 - \alpha \right) C^2 - 3 - 3
\alpha\right\} }
{l^2 \left( C^2 + 1 \right)^2} \nn
& + B \left[ X + \frac{2 C^4 \left\{ - 8 C^6 + \left( - 4 + \alpha \right) C^4
+ \left( 6 + 5 \alpha \right) C^2
+ 10 + 5 \alpha \right\}}{\left( C^2 + 1 \right)^3} \right] \, .
\end{align}
The Lagrangian density (\ref{NEM13cc}) still includes a parameter $Q$, which
appears as a solution in
(\ref{NEM5}) but as clear in (\ref{NEM6}), the parameter $Q$ can be always
absorbed into the
redefinition of $A_\mu$ as$A_\mu \to Q^2 A_\mu$.
Therefore we can find that this model given by the Lagrangian density
(\ref{NEM13cc}) has
only two coupling constants $\alpha$ and $l^2$.
In terms of $\alpha$ and $l^2$, the horizon radii $r_1$ and $r_2$ are given by
\begin{equation}
\label{r12}
r_1 = l \sqrt{\frac{\alpha - \sqrt{\alpha^2 - 4}}{2}} \, , \quad
r_2 = l \sqrt{\frac{\alpha + \sqrt{\alpha^2 - 4}}{2}} \, .
\end{equation}
This tells that by keeping the coupling constant $\alpha$ and $l^2$ to be
constant,
we cannot consider the Nariai limit, where $r_1 \to r_2$, or, $\alpha \to 2$.
In order to solve this problem, we make the parameter $\alpha$ a dynamical
field and we add the following Lagrangian density,
\begin{equation}
\label{Balpha}
\mathcal{L}_\mathrm{B\alpha} = H^\mu \partial_\mu \alpha \, .
\end{equation}
Then by the variation of $H^\mu$, we obtain the equation,
\begin{equation}
\label{Balpha2}
0=\partial_\mu \alpha \, ,
\end{equation}
and therefore $\alpha$ becomes a constant, whose value could be dynamically
determined.
Due to the equation (\ref{Balpha2}), the energy momentum tensor coming from the
Lagrangian
(\ref{Balpha}) vanishes and therefore the Lagrangian does not contribute to the
geometry.
Then we can consider the Nariai limit by making $\alpha \to 2$,
\begin{equation}
\label{Nrlim2}
r_1 = l - \epsilon\, , \quad r_2 = l + \epsilon \, , \quad
r = l + \epsilon \sin \theta \, , \quad t = \frac{\tau}{\epsilon}\, ,
\end{equation}
and taking the limit of $\epsilon \to 0$.
By the definition of (\ref{Nrlim2}), $\e^{2\nu(r)}$ in (\ref{NEM12B}) behaves
as
\begin{equation}
\label{Nrlim3B}
\e^{2\nu(r)} \to - \frac{2\epsilon^2}{l^2} \cos^2 \theta \, ,
\end{equation}
and therefore the metric is given by
\begin{equation}
\label{Nrlim4B}
ds^2 = \frac{2\cos^2 \theta}{l^2} d\tau^2 - \frac{l^2}{2} d\theta^2
+ l^2 \sum_{i,j=1}^2 \tilde g_{ij} dx^i dx^j\, .
\end{equation}
Further redefining
\begin{equation}
\label{Nrlim5}
\cos \theta = \frac{1}{\cosh \frac{2\rho}{l^2}} \, , \quad
\mbox{that is} \quad
d\theta = \frac{2}{l^2 \cosh \frac{2\rho}{l^2}} d\rho \, ,
\ee
we obtain the following metric
\begin{equation}
\label{Nrlim6}
ds^2 = \frac{2}{l^2\cosh^2 \frac{\rho}{l^2}} \left(d\tau^2 - d\rho^2 \right)
+ l^2 \sum_{i,j=1}^2 \tilde g_{ij} dx^i dx^j\, ,
\end{equation}
which is similar to the metric in the Nariai space-time.

The energy density (\ref{NEM13}) includes the contribution from the cosmological
constant because we are considering the asymptotically anti-de Sitter space-time.
By subtracting the contribution from the cosmological constant, we find
\begin{equation}
\label{NemA1}
\rho_\mathrm{EM} = \rho + \frac{6}{r_1 r_2}
= \frac{ 2 \left\{ \left( 2 r_1 r_2 + r_1^2 + r_2^2 \right) r^2
  - 3 \left( r_1^3 r_2 + r_1 r_2^3\right)\right\} }{r_1 r_2 \left( r^2 + r_1 r_2 \right)^2} \, .
\end{equation}
Therefore in the region
\begin{equation}
\label{NemA2}
r < \sqrt{\frac{3 \left( r_1^3 r_2 + r_1 r_2^3\right)}{2 r_1 r_2 + r_1^2 + r_2^2}} \, ,
\end{equation}
the energy density $\rho_\mathrm{EM}$ becomes negative and therefore the Weak and
Dominant Energy Conditions are not satisfied although the Null Energy Condition
is always satisfied, what is clear from (\ref{NNem14}).
Although we need to subtract the contributions from the cosmological constant,
there could be a region where the Strong Energy Condition is not 
satisfied.

\subsection{Regular Black Hole with Three Horizons}

We may also consider another example,
\begin{equation}
\label{NEM16B}
\e^{2\nu(r)} = - \frac{(r^2 - r_1^2)(r^2 - r_2^2)(r^2 - r_3^2)}
{l^2 \left( r^2 +\frac{r_1 r_2 r_3}{l}\right)^2} \, .
\end{equation}
In the limit $r\to \infty$, we find $\e^{2\nu(r)} \to - \frac{r^2}{l^2}$.
Therefore the space-time is the de Sitter space-time.
Then by using (\ref{NEM7}), we find
\begin{align}
\label{NEM17}
\rho =& - \frac{2}{l^2 r^2\left( r^2 +\frac{r_1 r_2 r_3}{l}\right)^3 } \nn
& \times \left[ - 3 r^8 + \left( r_1^2 + r_2^2 + r_3^2+ l^2
  - \frac{7 r_1 r_2 r_3}{l} \right) r^6
+ \left( r_2^2 r_3^2 + r_3^2 r_1^2 + r_1^2 r_2^2 + 3 l r_1 r_2 r_3
+ \frac{5 r_1 r_2 r_3}{l} \left(r_1^2 + r_2^2 + r_3^3 \right) \right) r^4
\right. \nn
& \left. \qquad - \frac{3 r_1 r_2 r_3}{l} \left( r_2^2 r_3^2
+ r_3^2 r_1^2 + r_1^2 r_2^2 \right) r^2
+ \frac{2 r_1^3 r_2^3 r_3^3}{l} \right] \, .
\end{align}
Now Eq.~(\ref{NEM9}) has the following form,
\begin{align}
\label{NEM18}
X =& \frac{2}{l^2 r^2\left( r^2 +\frac{r_1 r_2 r_3}{l}\right)^4 } \nn
& \times \left[ \left( - l^2 - r_1^2 - r_2^2 - r_3^2 - \frac{2 r_1 r_2 r_3}{l}
\right) r^8
\right. \nn
& + \left( - 2 \left( r_2^2 r_3^2 + r_3^2 r_1^2 + r_1^2 r_2^2 \right)
  - \frac{8 r_1 r_2 r_3}{l} \left( r_1^2 + r_2^2 + r_3^2 \right) - 4 r_1 r_2
r_3 l
  - \frac{14 r_1^3 r_2^3 r_3^3}{l^2} \right) r^6 \nn
& \left. + \left( \frac{5 r_1^2 r_2^2 r_3^2}{l^2} \left( r_1^2 + r_2^2 + r_3^2
\right)
  + \frac{10 r_1 r_2 r_3}{l} \left( r_2^2 r_3^2 + r_3^2 r_1^2 + r_1^2 r_2^2
\right)
+ 3 r_1^2 r_2^2 r_3^2 \right) r^4 - \frac{8 r_1^3 r_2^3 r_3^3}{l} r^2
  - \frac{2 r_1^4 r_2^4 r_3^4}{l^2} \right] \, .
\end{align}
In order to obtain the Lagrangian density $\mathcal{L}$, we need to solve
Eq.~(\ref{NEM18})
with respect to $r$ but the equation is the 5th order algebraic equation with
respect to $r$, which
cannot be solved algebraically.
Furthermore the one-to-one correspondence between $X$ and $r$ is not satisfied
in general and
therefore we cannot construct the Lagrangian density (\ref{NEM8}) in an
explicit form.
Then one may use the Lagrangian density in (\ref{LBCX}), and obtain
\begin{align}
\label{LBCX2B}
\mathcal{L}_{BCX} =& \frac{X}{C^2}
+ \frac{2}{l^2 C^2\left( C^2 +\frac{r_1 r_2 r_3}{l}\right)^3 }
\left[ - 3 C^8 + \left( r_1^2 + r_2^2 + r_3^2+ l^2
  - \frac{7 r_1 r_2 r_3}{l} \right) C^6 \right. \nn
& + \left( r_2^2 r_3^2 + r_3^2 r_1^2 + r_1^2 r_2^2 + 3 l r_1 r_2 r_3
+ \frac{5 r_1 r_2 r_3}{l} \left(r_1^2 + r_2^2 + r_3^3 \right) \right) C^4 \nn
& \left. \qquad - \frac{3 r_1 r_2 r_3}{l} \left( r_2^2 r_3^2
+ r_3^2 r_1^2 + r_1^2 r_2^2 \right) C^2
+ \frac{2 r_1^3 r_2^3 r_3^3}{l} \right] \nn
& + B \left[ X - \frac{2}{l^2 C^2\left( C^2 +\frac{r_1 r_2 r_3}{l}\right)^4 }
\right. \nn
& \times \left\{ \left( - l^2 - r_1^2 - r_2^2 - r_3^2 - \frac{2 r_1 r_2 r_3}{l}
\right) C^8
\right. \nn
& + \left( - 2 \left( r_2^2 r_3^2 + r_3^2 r_1^2 + r_1^2 r_2^2 \right)
  - \frac{8 r_1 r_2 r_3}{l} \left( r_1^2 + r_2^2 + r_3^2 \right) - 4 r_1 r_2
r_3 l
  - \frac{14 r_1^3 r_2^3 r_3^3}{l^2} \right) C^6 \nn
& + \left( \frac{5 r_1^2 r_2^2 r_3^2}{l^2} \left( r_1^2 + r_2^2 + r_3^2 \right)
  + \frac{10 r_1 r_2 r_3}{l} \left( r_2^2 r_3^2 + r_3^2 r_1^2 + r_1^2 r_2^2
\right)
+ 3 r_1^2 r_2^2 r_3^2 \right) C^4 \nn
& \left. \left. - \frac{8 r_1^3 r_2^3 r_3^3}{l} C^2
  - \frac{2 r_1^4 r_2^4 r_3^4}{l^2} \right\} \right] \, .
\end{align}
We now define parameters $\alpha$, $\beta$, and $\gamma$ as follows,
\begin{align}
\label{NEM19B}
& C^2 \to \frac{r_1 r_2 r_3}{l}C^2\, , \quad X \to \frac{r_1 r_2 r_3}{l} X \, ,
\quad
B \to \frac{l}{r_1 r_2 r_3} B \, , \nn
&\alpha \equiv \frac{l \left( r_1^2 + r_2^2 + r_3^2+ l^2 \right)}{r_1 r_2 r_3}
\, , \quad
\beta \equiv \frac{l^2 \left( r_2^2 r_3^2 + r_3^2 r_1^2 + r_1^2 r_2^2 \right)}
{r_1^2 r_2^2 r_3^2} \, , \quad
\gamma \equiv \frac{l^3}{r_1 r_2r_3} \, .
\end{align}
Then the Lagrangian density (\ref{LBCX2B}) has the following form
\begin{align}
\label{LBCX3B}
\mathcal{L}_{BCX} =& \frac{X}{C^2}
+ \frac{ 2 \left\{ - 3 C^8 + \left( \alpha - 7 \right) C^6
+ \left( 5 \alpha + \beta - 2 \gamma \right) C^4 - 3 \beta C^2 + 2 \right\}}
{l^2 C^2\left( C^2 + 1 \right)^3 } \nn
& + B \left[ X - \frac{2 \left\{ \left( - \beta - 2 \right) C^8
+ \left( - 8 \alpha - 2 \beta + 4 \gamma - 14 \right) C^6
+ \left( 5 \alpha + 10 \beta - 2 \gamma \right) C^4
  - 8 C^2 -2\right\} }{l^2 C^2\left( C^2 +1 \right)^4 } \right] \, .
\end{align}
This tells that this model has four independent parameters $l$, $\alpha$,
$\beta$, and $\gamma$
besides the gravitational constant.
Therefore when we consider the Nariai limit, we need to keep these parameters
constant.
By assuming $r_2,r_3\to r_0$, we find
\begin{equation}
\label{NEM20B}
\alpha = \frac{l \left( r_1^2 + 2 r_0^2 + l^2 \right)}{r_1 r_0^2} \, , \quad
\beta = \frac{l^2 \left( r_0^2 + 2 r_1^2 \right)}
{r_1^2 r_0^2} \, , \quad
\gamma = \frac{l^3}{r_1 r_0^2} \, .
\end{equation}
Then by deleting $r_1$, we obtain the following equations,
\begin{equation}
\label{NEM21}
\alpha = \frac{\gamma}{l^2} \left( \frac{l^6}{\gamma^2 r_0^4} + 2 r_0^2 + l^2
\right) \, , \quad
\beta = \frac{\gamma^2}{l^4} r_0^4 + \frac{2l^2}{r_0^2}\, .
\end{equation}
Because the above two equations are independent with each other,
there is no solution for $r_0$ in general.
Therefore one cannot get the Nariai limit in general.
Then we may use the formulation as in (\ref{Balpha}).
Instead of using the formulation, we may consider a special case,
\begin{equation}
\label{special}
\alpha = \gamma + 3 \gamma^{\frac{1}{3}}\, , \quad \beta = 3
\gamma^{\frac{2}{3}} \, .
\end{equation}
Then one has a solution,
\begin{equation}
\label{special2}
r_0 = r_1 = \gamma^{-\frac{1}{3}} l \, ,
\end{equation}
that is, the radii of three horizons coincide with each other.
We may also consider the case that the parameters $\alpha$, $\beta$, and
$\gamma$ in
(\ref{NEM20B}) are given by two parameters $\xi_0$ and $\xi_1$ as follows,
\begin{equation}
\label{NEM21B}
\alpha = \frac{\xi_0^2}{\xi_1} + 2 \xi_1 + \xi_0^2 \xi_1\, , \quad
\beta = \xi_1^2 + 2 \xi_0^2 \, , \quad \gamma = \xi_0^2 \xi_1 \, .
\end{equation}
Then the solution for $r_0$ and $r_1$ is given by
\begin{equation}
\label{NEM21C}
r_0 = \frac{l}{\xi_0} \, , \quad r_1 = \frac{l}{\xi_1} \, .
\end{equation}
Hence even if we consider the infinitesimal shift from $r_2=r_3=r_0$,
\begin{equation}
\label{NEM21D}
r_2 = r_0 - \epsilon\, , \quad
r_3 = r_0 + \epsilon \, ,
\end{equation}
the parameters $\alpha$, $\beta$, and $\gamma$ do not change or more exactly
the variations of the parameters are $\mathcal{O}\left( \epsilon^2 \right)$.
Then as in (\ref{Nrlim2}), by redefining
\begin{equation}
\label{Nrlim2B}
r = r_0 + \epsilon \sin \theta \, , \quad t = \frac{\tau}{\epsilon}\, ,
\end{equation}
we obtain the Nariai limit,
\begin{equation}
\label{NEM21E}
\e^{2\nu(r)} \to
\frac{4 r_0^2 (r_0^2 - r_1^2) \epsilon^2 \cos^2 \theta}
{l^2 \left( 1 +\frac{r_1}{l}\right)^2} \, ,
\end{equation}
and the metric is given by
\begin{equation}
\label{NEM21F}
ds^2 = - \frac{4 r_0^2 (r_0^2 - r_1^2) \cos^2 \theta}
{l^2 \left( 1 +\frac{r_1}{l}\right)^2} d\tau^2
+ \frac{l^2 \left( 1 +\frac{r_1}{l}\right)^2}{4 r_0^2 (r_0^2 - r_1^2)}d\theta^2
+ r_0^2 \sum_{i,j=1}^2 \tilde g_{ij} dx^i dx^j\, .
\end{equation}

\section{$F(R)$ gravity with Non-Linear Electromagnetism \label{Sec3}}

In this section, we consider $F(R)$ theory instead of the Einstein gravity.
We show that the extension of above formulation is rather straightforward.
Note that the black holes in $F(R)$ gravity have been well-studied
\cite{Clifton:2011jh,Olmo:2006eh,Briscese:2007cd,delaCruzDombriz:2009et,
DeLaurentis:2012st,Sebastiani:2010kv,Olmo:2011ja}
including the solution of the black hole with multi-horizons in
$F(R)$-gravity,
(see \cite{Nojiri:2013su}). It is interesting that for such $F(R)$ black
holes
as was shown in Refs.~\cite{Nojiri:2013su,Nojiri:2014jqa,Addazi:2016hip}, the
anti-evaporation phenomena in the Nariai space-time
\cite{Bousso:1997wi,Nojiri:1998ue,Nojiri:1998ph} may occur.
In this section, we also prove that within only $F(R)$ gravity, we cannot
construct the regular
black hole solution of the Schwarzschild type (\ref{GBiv}).

\subsection{Regular Black Hole in $F(R)$ gravity with Non-Linear
Electromagnetism}

We start with the action of $F(R)$ gravity coupled with matter:
\be
\label{M1}
S = \int d^4 x \sqrt{-g}\left\{ F(R) + L_m\right\}\, .
\ee
Here $F(R)$ is a function of the scalar curvature and $L_m$ is a matter
Lagrangian.
The equation of the motion is given by
\begin{align}
\label{eqFR2}
0 = & - \frac{1}{2} \e^{2\nu} F(R) - \e^{4\nu} \left\{
\nu'' + 2 {\nu'}^2 + \frac{2\nu'}{r}\right\} F'(R)
  - \e^{4\nu}\nu' \frac{d F'(R)}{dr}
+ \frac{\e^{2 \nu}}{r^2}\frac{d}{dr}
\left( \e^{2\nu} r^2 \frac{d F'(R)}{dr} \right) - \frac{1}{2}\rho\e^{2\nu} \, ,
\\
\label{eqFR2B}
0 = & \frac{1}{2} \e^{- 2\nu} F(R) + \left\{ \nu'' + 2 {\nu'}^2 + \frac{2
\nu'}{r} \right\} F'(R)
+ \frac{d^2 F'(R)}{dr^2} + \nu' \frac{d F'(R)}{dr}
  - \frac{\e^{- 2 \nu}}{r^2}\frac{d}{dr}
\left( \e^{2\nu} r^2 \frac{d F'(R)}{dr} \right) + \frac{1}{2} p \e^{-2\nu}\, ,
\\
\label{eqFR2C}
0 =& \frac{r^2}{2} F(R)
  - \left\{ 1 + \left( - 1 - 2 r \nu' \right)\e^{2\nu}\right\} F'(R)
+ \e^{2\nu} r \frac{d F'(R)}{dr} - \frac{d}{dr}
\left( \e^{2\nu} r^2 \frac{d F'(R)}{dr} \right) + \frac{r^2}{2} p_T \, .
\end{align}
Even in $F(R)$ gravity theory, we have Eqs.~(\ref{NEM8}) and (\ref{NEM9}),
again but
Eq.~(\ref{NEM7}) is replaced by (\ref{eqFR2}).
For the geometry given by $\e^{2\nu(r)}$ in (\ref{GBiv}),
Eq.~ (\ref{eqFR2}) determines the $r$-dependence of $\rho$, $\rho=\rho(r)$.
By solving (\ref{NEM9}), we find the $r$-dependence of $X$, which can be
solved with respect to $r$, $X=X(r)$.
Substituting the expression $X(r)$ into (\ref{NEM8}), we obtain the Lagrangian
density $\mathcal{L}$.
Even if one cannot solve (\ref{NEM9}), we can use the Lagrangian density
(\ref{LBCX}) by using the auxiliary fields $B$ and $C$.

For the example (\ref{NEM12B}), one finds
\begin{equation}
\label{NEM12BB}
R = \frac{ 2 \left\{ - 6 r^6 + \left( r_1^2 + r_2^2 - 16 r_1 r_2 \right) r^4
+ \left( 3 r_1^3 r_2 + 3 r_1 r_2^3 - 12 r_1^2 r_2^2 \right) r^2
+ 6 r_1^4 r_2^2 + 6 r_1^2 r_2^4 + 6 r_1^3 r_2^3 \right\} }
{ r_1 r_2 \left( r^2 + r_1 r_2 \right)^3} \, .
\end{equation}
Let us consider the following $F(R)$ gravity,
\be
\label{FRR1}
F(R)= \frac{R}{2} +\frac{c}{2} R^2\, ,
\ee
with a constant $c$.
Using (\ref{eqFR2}), we find
\begin{align}
\label{NEM13RR}
\rho =& \frac{ 2 \left\{ 3r^4 + \left( 4 r_1 r_2 - r_1^2 - r_2^2 \right) r^2 -
3 r_1^3 r_2^3
  - 3 r_1^3 r_2 - 3 r_1 r_2^3\right\} }{r_1 r_2 \left( r^2 + r_1 r_2 \right)^2}
\nn
& + \frac{4c}{r_1^2 r_2^2 \left( r^2 + r_1 r_2 \right)^6} \left[ 72 r^{12}
+ \left( 408 r_1 r_2 - 12 r_1^2 - 12 r_2^2 \right) r^{10}
+ \left( - 68 r_1^3 r_2 - 68 r_1 r_2^3 - 3 r_1^4 - 3 r_2^4 - 950 r_1^2 r_2^2
\right) r^8
\right. \nn
& \qquad \qquad \qquad \qquad
+ \left( - 18 r_1^5 r_2 - 18 r_1 r_2^5 + 1308 r_1^3 r_2^3 - 84 r_1^4 r_2^2 -
84 r_1^2 r_2^4
\right) r^6 \nn
& \qquad \qquad \qquad \qquad
+ \left( - 207 r_1^6 r_2^2 - 207 r_1^2 r_2^4 - 1056 r_1^5 r_2^3 - 1056 r_1^3
r_2^5
  - 618 r_1^4 r_2^4 \right) r^4 \nn
& \qquad \qquad \qquad \qquad
+ \left( 192 r_1^7 r_2^3 + 192 r_1^3 r_2^7 + 184 r_1^6 r_2^4 + 184 r_1^4 r_2^6
+ 506 r_1^5 r_2^5 \right) r^2 \nn
& \left. \qquad \qquad \qquad \qquad
+ 72 r_1^8 r_2^4 + 72 r_1^4 r_2^8 - 36 r_1^7 r_2^5 - 36 r_1^5 r_2^7
  - 144 r_1^6 r_2^6 \right]\, .
\end{align}
Then by using (\ref{NEM9}), one gets
\begin{align}
\label{NEM13bRR}
X =& - \frac{2 r^4 \left\{ - 8 r^6 + \left( - 4 r_1 r_2 + r_1^2 + r_2^2 \right)
r^4
+ \left( 6 r_1^2 r_2^2 + 5 r_1^3 r_2 + 5 r_1 r_2^3 \right) r^2
+ 10 r_1^3 r_2^3 + 5 r_1^4 r_2^2 + 5 r_1^2 r_2^4 \right\}}{
r_1 r_2 \left( r^2 + r_1 r_2 \right)^3} \nn
& + \frac{ 8 c \left(r_1 + r_2 \right)^2 r^4}{r_1^2 r_2^2 \left( r^2 + r_1 r_2
\right)}
\left[ - 6 r^{10} + \left( -3 r_1^2 - 3 r_2^2 -32 r_1 r_2 \right) r^8
+ \left( - 21 r_1^3 r_2 - 21 r_1 r_2^3 + 52 r_1^2 r_2^2 \right) r^6 \right. \nn
& \qquad \qquad \qquad \qquad
+ \left( - 387 r_1^4 r_2^2 - 387 r_1^2 r_2^4 - 1212 r_1^3 r_2^3 \right) r^4
+ \left( 687 r_1^5 r_2^3 + 687 r_1^3 r_2^5 + 142 r_1^4 r_2^4 \right) r^2 \nn
& \left. \qquad \qquad \qquad \qquad
+ 120 r_1^6 r_2^4 + 120 r_1^4 r_2^6 - 440 r_1^5 r_2^5
\right]
\, .
\end{align}
By using the Lagrangian density in (\ref{LBCX}), we find
\begin{align}
\label{NEM13ccc}
\mathcal{L}_{BCX} =& \frac{X}{C^2}
  - \frac{ 2 \left\{ 3C^4 + \left( 4 r_1 r_2 - r_1^2 - r_2^2 \right) C^2 - 3
r_1^3 r_2^3
  - 3 r_1^3 r_2 - 3 r_1 r_2^3\right\} }{r_1 r_2 \left( C^2 + r_1 r_2 \right)^2}
\nn
& - \frac{4c}{r_1^2 r_2^2 \left( C^2 + r_1 r_2 \right)^6} \left[ 72 C^{12}
+ \left( 408 r_1 r_2 - 12 r_1^2 - 12 r_2^2 \right) C^{10}
+ \left( - 68 r_1^3 r_2 - 68 r_1 r_2^3 - 3 r_1^4 - 3 r_2^4 - 950 r_1^2 r_2^2
\right) C^8
\right. \nn
& \qquad \qquad \qquad \qquad
+ \left( - 18 r_1^5 r_2 - 18 r_1 r_2^5 + 1308 r_1^3 r_2^3 - 84 r_1^4 r_2^2 -
84 r_1^2 r_2^4
\right) C^6 \nn
& \qquad \qquad \qquad \qquad
+ \left( - 207 r_1^6 r_2^2 - 207 r_1^2 r_2^4 - 1056 r_1^5 r_2^3 - 1056 r_1^3
r_2^5
  - 618 r_1^4 r_2^4 \right) C^4 \nn
& \qquad \qquad \qquad \qquad
+ \left( 192 r_1^7 r_2^3 + 192 r_1^3 r_2^7 + 184 r_1^6 r_2^4 + 184 r_1^4 r_2^6
+ 506 r_1^5 r_2^5 \right) C^2 \nn
& \left. \qquad \qquad \qquad \qquad
+ 72 r_1^8 r_2^4 + 72 r_1^4 r_2^8 - 36 r_1^7 r_2^5 - 36 r_1^5 r_2^7
  - 144 r_1^6 r_2^6 \right]
\nn
&+ B \left[ X
+ \frac{2 C^4 \left\{ - 8 C^6 + \left( - 4 r_1 r_2 + r_1^2 + r_2^2 \right) C^4
+ \left( 6 r_1^2 r_2^2 + 5 r_1^3 r_2 + 5 r_1 r_2^3 \right) C^2
+ 10 r_1^3 r_2^3 + 5 r_1^4 r_2^2 + 5 r_1^2 r_2^4 \right\}}{
r_1 r_2 \left( C^2 + r_1 r_2 \right)^3} \right. \nn
& - \frac{ 8 c \left(r_1 + r_2 \right)^2 C^4}{r_1^2 r_2^2 \left( C^2 + r_1 r_2
\right)}
\left[ - 6 r^{10} + \left( -3 r_1^2 - 3 r_2^2 -32 r_1 r_2 \right) C^8
+ \left( - 21 r_1^3 r_2 - 21 r_1 r_2^3 + 52 r_1^2 r_2^2 \right) C^6 \right. \nn
& \qquad \qquad \qquad \qquad
+ \left( - 387 r_1^4 r_2^2 - 387 r_1^2 r_2^4 - 1212 r_1^3 r_2^3 \right) C^4
+ \left( 687 r_1^5 r_2^3 + 687 r_1^3 r_2^5 + 142 r_1^4 r_2^4 \right) C^2 \nn
& \left. \left. \qquad \qquad \qquad \qquad
+ 120 r_1^6 r_2^4 + 120 r_1^4 r_2^6 - 440 r_1^5 r_2^5 \right]\right] \, .
\end{align}
Similarly, we can calculate the Lagrange density corresponding to the geometry
(\ref{NEM16B}) although the corresponding expression is very complicated.

With several horizons, one can consider the limit where the radius of one
horizon coincides with
that of another horizon, which is called the Nariai limit and the obtained
space-time is the Nariai
space-time. Usually by the Hawking radiation, the radius of the horizon
decreases but in case of the
Nariai space-time, the radius can increase by including the quantum effects,
which is called as
the anti-evaporation \cite{Bousso:1997wi,Nojiri:1998ue,Nojiri:1998ph}.
In \cite{Nojiri:2013su,Nojiri:2014jqa,Addazi:2016hip}, the anti-evaporation in $F(R)$
gravity was investigated and it was shown that
the anti-evaporation can occur even at the classical level.
The corresponding study for the above regular black hole will be done
elsewhere.

\subsection{A No-Go Theorem in $F(R)$ Gravity}

It is found above that the solutions describing the non-singular black
hole with multi-horisons can be obtained in $F(R)$ gravity coupled with the
non-linear electromagnetic field.
It is interesting to understand if we can realize the non-singular black holes
only by $F(R)$ gravity without the
account of non-linear electrodynamics. For this purpose, we consider the case
of vacuum, where $\rho=p=p_T=0$.
Combining Eq.~(\ref{eqFR2}) and Eq.~(\ref{eqFR2B}) we find
\be
\label{eqFR3}
0 = \frac{d^2 F'(R)}{dr^2} \, ,
\ee
that is,
\be
\label{eqFR4}
F'(R) = f_0 + f_1 r \, ,
\ee
with constants $f_0$ and $f_1$. Hence, one can rewrite the equations in
(\ref{eqFR2}) as follows,
\begin{align}
\label{eqFR5}
0 = & \frac{1}{2} F(R) + \left\{ \nu'' + 2 {\nu'}^2 + \frac{2 \nu'}{r} \right\}
\e^{2\nu}
\left( f_0 + f_1 r \right) - f_1 \nu' \e^{2\nu} - \frac{2 f_1 \e^{2\nu}}{r}
\, , \nn
0 =& \frac{1}{2} F(R)
  - \frac{1}{r^2}\left\{ 1 + \left( - 1 - 2 r \nu' \right)\e^{2\nu}\right\}
\left( f_0 + f_1 r \right)
  - \frac{f_1 \e^{2\nu}}{r} - 2 f_1 \nu' \e^{2\nu} \, .
\end{align}
Deleting $F(R)$ in (\ref{eqFR5}), we obtain,
\be
\label{eqFR6}
0 = \left\{ \left( \nu'' + 2 {\nu'}^2 - \frac{1}{r^2} \right) \e^{2\nu} +
\frac{1}{r^2}
\right\} \left( f_0 + f_1 r \right) + f_1 \nu' \e^{2\nu} - \frac{f_1
\e^{2\nu}}{r} \, ,
\ee
which is a differential equation for $\nu$.
By defining $N\equiv \e^{2\nu}$, we rewrite Eq.~(\ref{eqFR6}) as follows,
\begin{equation}
\label{eqFR7}
0 = \frac{f_0+ f_1 r}{2} N'' + \frac{f_1}{2} N' - \left( \frac{f_0}{r^2} +
\frac{2f_1}{r} \right) N
+ \frac{f_0 + f_1 r}{r^2} \, ,
\end{equation}
As a special case that $f_1=0$, the solution of (\ref{eqFR7}) is given by
\be
\label{eqFR8}
N=N^{(0)}(r) \equiv 1 + A r^2 + \frac{B}{r} \, ,
\ee
with constants $A$ and $B$. This is nothing but the (A)dS-Schwarzschild
solution.
On the other hand, in case that $f_0=0$, we find
\be
\label{eqFR9}
N = N^{(1)}(r) \equiv \frac{1}{2} + C^+ r^2 + \frac{C^-}{r^2} \, .
\ee
For (\ref{eqFR9}), we get the following scalar curvature
\be
\label{eqFR10}
R = \e^{2\nu} \left[ - \frac{d^2 \e^{2\nu}}{dr^2} - \frac{4}{r}
\frac{d\e^{2\nu}}{dr}
+ \frac{2\e^{-2\nu} - 2}{r^2} \right]
= \left( \frac{1}{2} + C^+ r^2 + \frac{C^-}{r^2} \right)
\left( - 8 C^+ - \frac{4 C^-}{r^4} - \frac{1}{r^2}\right) \, ,
\ee
which is singular at $r=0$.
In order to solve Eq.~(\ref{eqFR7}) for general case, we first consider the
following homogeneous differential equation,
\begin{equation}
\label{eqFR11}
0= \frac{f_0+f_1 r}{2} {\tilde N}'' + \frac{f_1}{2} {\tilde N}'
  - \left( \frac{f_0}{r^2} + \frac{2f_1}{r} \right) \tilde N \, ,
\end{equation}
whose trivial solution is $\tilde N = r^2$.
Then by assuming $\tilde N = A(r) r^2$, one finds the following equation,
\begin{equation}
\label{eqFR13}
0 = A'' + \left( \frac{2}{r} + \frac{2f_1}{f_0 + f_1 r} \right) A' \, ,
\end{equation}
whose solution is given by
\begin{equation}
\label{eqFR14}
A' = \frac{C_1}{r^2 \left(\frac{f_0 }{f_1} + r \right)^2}
= \frac{ C_1 f_1^2}{f_0^2} \left( \frac{1}{r} - \frac{1}{\frac{f_0 }{f_1} + r}
\right)^2
= \frac{ C_1 f_1^2}{f_0^2} \left( \frac{1}{r^2} - \frac{1}{\left( \frac{f_0
}{f_1} + r\right)^2}
  - \frac{2f_1}{f_0} \left( \frac{1}{r} - \frac{1}{\frac{f_0 }{f_1} + r} \right)
\right) \, ,
\end{equation}
with a constant $C_1$ of integration.
The general solution of $\tilde N$ is given by
\begin{equation}
\label{eqFR15}
\tilde N = C_1 \left( - \frac{f_1}{f_0}
+ \frac{1}{\frac{f_0 }{f_1} + r}
+ \frac{2f_1^3 r^2}{f_0^3} \ln \left( \frac{f_0 }{f_1 r} + 1 \right) \right)
+ C_2 r^2 \, ,
\end{equation}
with another constant $C_2$ of the integration.
This tells the expression of the general solution of $N$ as follows,
\begin{align}
\label{eqFR16}
N =& C_1 \left( - \frac{f_1}{f_0}
+ \frac{1}{\frac{f_0 }{f_1} + r}
+ \frac{2f_1^3 r^2}{f_0^3} \ln \left( \frac{f_0 }{f_1 r} + 1 \right) \right)
+ C_2 r^2 \nn
& - \frac{1}{2} + \frac{2f_1 r }{f_0}
+ \frac{2f_1^2 r^2}{f_0^2} \ln \left( \frac{f_0 }{f_1} + r \right)
+ \frac{2f_1 r}{f_0} \ln r + \frac{2f_1 r}{f_0}
  - \frac{2f_1 r}{f_0\left( \frac{f_0 }{f_1} + r \right)} \ln r
  - \frac{2f_1^2 r^2}{f_0^2} \left( \ln r - \ln \left( \frac{f_0 }{f_1} +
r\right) \right) \nn
& + \frac{4 f_1^2 r^2}{f_0^2} \int^r dr' \ln r' \left( \frac{1}{r'}
  - \frac{1}{\frac{f_0 }{f_1} + r'} \right) \, .
\end{align}
In order to avoid the singularity at $r=0$, we require
$N=\e^{2\nu} = 1 + \mathcal{O} \left(r^2\right)$ but the expression in
(\ref{eqFR16})
is singular if $C_1\neq 0$.
Even if $C_1 =0$, we find $N=\e^{2\nu} \to\frac{1}{2}$ in the limit of $r\to 0$
and therefore there
remains a singularity.
Therefore, we cannot realize the non-singular black hole if we assume the
metric is given in the
form of the Schwarzschild type (\ref{GBiv}).

\section{Thermodynamics of Regular Multi-Horizon Black Holes \label{Sec4}}

Let us study the thermodynamics of the obtained regular black hole.
The Hawking temperature $T$ is now defined by
\begin{equation}
\label{HT}
T = \frac{1}{4\pi} \left. \frac{d \e^{2\nu}}{dr} \right|_{r=r_\mathrm{H}}\, .
\end{equation}
Here $r_\mathrm{H}$ is the radius of the horizon.

First we consider the metric (\ref{NNem1}).
Then when $r_\mathrm{H}=r_2>r_1$ $\left(r_\mathrm{H}=r_1\right)$,
we find the temperature $T=T_2$ $\left( T_1 \right)$,
\begin{equation}
\label{NNem7}
4\pi T_2 =  \frac{ \left(r_2 + r_0\right) \left( r_2 - r_1 \right) }{\beta + r_2^3}\quad
\left(  4\pi T_1 = \frac{ \left(r_1 + r_0\right) \left( r_1 - r_2 \right) }{\beta + r_1^3}<0 \right) \, .
\end{equation}
By solving (\ref{NNem5}) with respect to $r_1$, one gets
\begin{equation}
\label{NNem8}
r_1 = \frac{ \left( - 1 \pm \sqrt{ 4 \tilde \alpha r_2^2 - 3} \right) r_2}
{2 \left( 1 - \tilde \alpha r_2^2 \right)} \, .
\end{equation}
As given in (\ref{NNem6}), we are assuming $\tilde \alpha r_2^2 > 3$ and therefore
the expression (\ref{NNem8}) is real and because $r_1>0$, we find
\begin{equation}
\label{NNem9}
r_1 = \frac{ \left( 1 + \sqrt{ 4 \tilde \alpha r_2^2 - 3} \right) r_2}
{2 \left( \tilde \alpha r_2^2 - 1 \right)} \, .
\end{equation}
Therefore Eq.~(\ref{NNem2}) shows
\begin{equation}
\label{NNem10}
r_0 = \frac{ \left( 1 + \sqrt{ 4 \tilde \alpha r_2^2 - 3} \right) r_2}
{ 2 \tilde \alpha r_2^2 - 1 + \sqrt{ 4 \tilde \alpha r_2^2 - 3}}\, , \quad
\beta = \frac{ \left( 1 + \sqrt{ 4 \tilde \alpha r_2^2 - 3} \right)^2 r_2^3}
{ 2 \left( \tilde \alpha r_2^2 - 1 \right)
\left(2 \tilde \alpha r_2^2 - 1 + \sqrt{ 4 \tilde \alpha r_2^2 - 3}\right)}\, ,
\end{equation}
and
\begin{equation}
\label{NNem11}
4\pi T_2 = \frac{2 \left( \tilde \alpha r_2^2 + \sqrt{ 4 \tilde \alpha r_2^2 - 3} \right)
\left( 2 \tilde \alpha r_2^2 - 3 - \sqrt{ 4 \tilde \alpha r_2^2 - 3} \right)}
{\left(  4 {\tilde \alpha}^2 r_2^4 - 1
+  2 \tilde \alpha r_2^2 \sqrt{ 4 \tilde \alpha r_2^2 - 3}\right) r_2}\, .
\end{equation}
In the extremal limit (\ref{NNem6}), the temperature $T_2$ vanishes.
On the other hand, for large $r_2$, $T_2$ behaves as
\begin{equation}
\label{NNem11}
4\pi T_2 \sim \frac{1}{r_2}\, .
\end{equation}
As in the usual black hole, the large black hole has low temperature.
For the large black hole, the entropy $\mathcal{S}$ is given by
\begin{equation}
\label{NNem12}
\mathcal{S}= \frac{A}{4} = \pi r_2^2
\sim \frac{1}{16 \pi T_2^2} \, .
\end{equation}
By using the thermodynamical relation, $dE = T d\mathcal{S}$, we may estimate
the thermodynamical energy.
\begin{equation}
\label{NNem13}
E \sim \frac{1}{8\pi T_2} = \frac{r_2}{2} + E_0\, ,
\end{equation}
which coincides with the ADM mass $M= \frac{r_2}{2}$ if we choose the constant $E_0$
of the integration to vanish.

Then for the metric (\ref{NEM12B}), the temperature is given by
\begin{equation}
\label{NEM12TH}
T = T_2 = \frac{ r_2 - r_1 }{2\pi r_1 r_2}
= \frac{r_2 - \frac{l^2}{r_2} }{2\pi l^2}
\, .
\end{equation}
Here we have evaluated the temperature at the outer horizon $r=r_2> r_1$.
The temperature $T$ vanishes in the extremal limit where $r_2\to r_1>0$ as for
the standard singular black holes with multi-horizons like the
Reissner-Nordstr\" om black hole.
At the inner horizon at $r=r_1$, we obtain the expression by exchanging $r_1$
and $r_2$ although the
obtained expression is negative,
\begin{equation}
\label{NEM12TH2}
T = T_1 = - \frac{r_2 - r_1}{2\pi r_1 r_2} \, .
\end{equation}
The negativity could be understood if the temperature measured in the region
$r<r_1$ could be given by
replaceing $\frac{d}{dr}$ with $- \frac{d}{dr}$ in the definition of
(\ref{HT}), which is positive $T=-T_1 = T_2>0$.
In case of Einstein gravity, the entropy $\mathcal{S}$ on the outer horizon is
given by
\begin{equation}
\label{ent}
\mathcal{S}= \frac{A}{4} = \pi r_2^2\, .
\end{equation}
In case of $F(R)$ gravity (\ref{FRR1}), there is a correction,
\begin{equation}
\label{entFR}
\mathcal{S}= \frac{A}{4}\left. F'(R)\right|_{r=r_2}
= \pi r_2^2 \left( 1 + 4c
\frac{ - 5 r_2^4 - 13 r_1 r_2^3 - 5 r_1^2 r_2^2 + 9 r_1^3 r_2 + 6 r_1^4 r_2}
{r_1 \left( r_1 + r_2 \right)} \right) \, .
\end{equation}
Here $r_1$ is given by
\begin{equation}
\label{r1}
r_1 = \frac{l^2}{r_2} \, .
\end{equation}
By definition in (\ref{ent}) or (\ref{entFR}), the entropy $\mathcal{S}$ is
always positive for
$F(R)$ gravity as long as we do not consider the anti-gravity region, where
$F'(R)<0$ and therefore
this region is unphysical.

We now consider the thermodynamics by fixing the parameter $l^2$ but varying
the parameter $\alpha$.
Eq.~(\ref{NEM12TH}) can be solved with respect to $r_2$ as follows,
\begin{equation}
\label{Tr2}
r_2 = \pi l^2 T \pm \sqrt{ \pi^2 l^4 T^2 + l^2}\, ,
\end{equation}
Because $r_2>0$, we choose the plus sign in (\ref{Tr2}),
\begin{equation}
\label{Tr2B}
r_2 = \pi l^2 T + \sqrt{ \pi^2 l^4 T^2 + l^2}\, ,
\end{equation}
We should note that when the temperature becomes large $T\to \infty$, the
radius of the outer
horizon becomes also large, whose situation is different from that in the
standard black hole, where the large temperature corresponds to the small
horizon radius.
Then the entropy (\ref{ent}) is expressed as
\begin{equation}
\label{ent3}
\mathcal{S} = \frac{\pi}{4} \left(
2 \pi^2 l^4 T^2 + l^2 + 2 \pi l^2 T \sqrt{ \pi^2 l^4 T^2 + l^2} \right) \, .
\end{equation}
We should note that the entropy does not vanish even if we consider the limit
of $T\to 0$, which corresponds to the Nariai limit $r_2\to r_1$,
\begin{equation}
\label{ent3B}
\mathcal{S} \to \frac{\pi l^2}{4} \, ,
\end{equation}
which is different from the standard thermodynamics.
This may indicate that there remain some information in the regular part at the
origin.
In case of $F(R)$ gravity (\ref{entFR}), the expression is rather complicated.
By using the thermodynamical relation, $dE = T d\mathcal{S}$, we may estimate
the thermodynamical energy.
\begin{equation}
\label{ent4}
E = \frac{\pi^3}{3} l^4 T^3
+ \frac{1}{l^2} \left(\pi^2 l^4 T^2 + l^2\right)^{\frac{3}{2} }
  - \frac{1}{2} \sqrt{ \pi^2 l^4 T^2 + l^2} + E_0 \, .
\end{equation}
Here $E_0$ is a constant of the integration.
In case of the AdS-Schwarzschild black hole, we find
\begin{equation}
\label{AdSS}
\e^{2\nu} = \frac{r^2}{l^2} + 1 - \frac{2M}{r} \, ,
\end{equation}
Here $M$ corresponds to the ADM mass of the black hole \cite{Arnowitt:1959ah}.
In case of Einstein gravity, the mass coincides with the thermodynamical energy
$E$.
On the other hand, in case of (\ref{NEM12B}), in the limit of $r\to \infty$,
we find
\begin{equation}
\label{rAdSS}
\e^{2\nu} \to \frac{r^2}{r_1 r_2} - \frac{r_1^2 + r_2^2 + r_1 r_2}{r_1 r_2}
+ \mathcal{O} \left( r^{-2} \right) \, .
\end{equation}
Then there is no parameter corresponding to the mass.
The discrepancy may occur due to the energy coming from the non-linear
electromagnetic field.
For large $T$, the thermodynamical energy $E$ in (\ref{ent4}) behaves as
\begin{equation}
\label{ent55}
E \sim \frac{4}{3} \pi^3 l^4 T^3 \, .
\end{equation}
Therefore the specific heat $C=\frac{dE}{dT}$ is positive, which is different
from the case of the
standard black hole, where the specific heat is negative.
This could be related with the observation in (\ref{Tr2B}), where the large
temperature
corresponds to the large horizon radius.

\section{Einstein-Gauss-Bonnet Gravity in Five Dimensions \label{Sec5}}

Let us now consider the Einstein-Gauss-Bonnet gravity in five-dimensions. It is
known that such theory is very useful in the study of the AdS/CFT
correspondence.

\subsection{General Formulation}

The action of Einstein-Gauss-Bonnet gravity coupled with matter is expressed by
\begin{equation}
\label{vi}
S=\int d^5 x \sqrt{-g}\left\{ \frac{R}{2} + c \left( R^2 -4 R_{\mu\nu}
R^{\mu\nu}
+ R_{\mu\nu\xi\sigma} R^{\mu\nu\xi\sigma} \right) + L_m \right\}\ .
\end{equation}
The equations of motion have the form
\begin{align}
\label{R3}
0=& \frac{1}{2}g_{\mu\nu}\left\{ \frac{R}{2} + c\left( R^2 - 4 R_{\rho\sigma}
R^{\rho\sigma}
+ R_{\rho\tau\xi\sigma} R^{\rho\tau\xi\sigma}\right) \right\} \nn
& - \frac{1}{2}R_{\mu\nu} + c\left(-2RR_{\mu\nu} + 4 R_{\mu\rho} R_\nu^{\
\rho}
+ 4 R_{\mu\ \nu}^{\ \rho\ \sigma} R_{\rho\sigma}
  - 2 R_\mu^{\ \rho\sigma\tau}R_{\nu\rho\sigma\tau} \right) + T_{\mu\nu} \ .
\end{align}
By assuming a spherically-symmetric static background as (\ref{GBiv}) but in 5
dimensions
\begin{equation}
\label{GBiv5d}
ds^2 = - \e^{2\nu (r)} dt^2 + \e^{-2\nu (r)} dr^2
+ r^2 \sum_{i,j=1}^3 \tilde g_{ij} dx^i dx^j\, ,
\end{equation}
the $(\mu,\nu)=(t,t)$, $(r,r)$ and $(i,j)$ components of Eq.~(\ref{R3}) have
the following expressions,
\begin{align}
\label{GBvid2B}
0=& - \frac{\e^{-2\nu}}{2}\left[ - \frac{24 c \e^{2\nu}
\left(1-\e^{2\nu}\right)\nu'}{r^3}
+ \frac{3\e^{2\nu}}{2}\left\{ - \frac{2\nu'}{r} + \frac{2\left(1-\e^{2\nu}\right)
\e^{-2\nu}}{r^2} \right\} \right] - \rho \e^{-2\nu} \, , \\
\label{GBviid2B}
0=& \frac{\e^{2\nu}}{2}\left[ - \frac{24 c \e^{2\nu}\left(1-\e^{2\nu}\right)
\nu'}{r^3}
+ \frac{3\e^{2\nu}}{2} \left\{- \frac{2\nu'}{r}
+ \frac{2\left(1-\e^{2\nu}\right) \e^{-2\nu}}{r^2} \right\} \right]
+ p \e^{2\nu}\ , \\
\label{GBviiid2B}
0=& \frac{1}{2r^2} \left[ -c\e^{2\nu}
\left\{ 2\left(1-\e^{2\nu}\right) \left( \frac{4 \left(\nu''
+ 2 {\nu'}^2\right)}{r^2}
+ \frac{16 \nu' }{r^3} \right)
  - \frac{16 {\nu'}^2}{r^2}\right\} \right. \nn
& \left. + \frac{\e^{2\nu}}{2}\left\{ - 2\nu''
  - 4 {\nu'}^2 - \frac{12\nu'}{r}
+ \frac{12 \left(1-\e^{2\nu}\right) \e^{-2\nu}}{r^2} \right\}
\right] + \frac{p_T}{r^2}\ .
\end{align}
We now consider the non-linear electromagnetic field as a matter as in
(\ref{NEM1}).
Then in five dimensions, instead of (\ref{NEM4}), we find
\begin{equation}
\label{NEM45d}
0 = \frac{d}{dr}\left( r^3 F^{0r} \frac{d \mathcal{L}}{dI} \right) \, .
\end{equation}
Using $I = -\frac{1}{2} \left( F_{0r} \right)^2$ as in (\ref{NEM5}), we find
$X \equiv Q \sqrt{ - 2I}$
\begin{equation}
\label{NEM55d}
r^3 \frac{d \mathcal{L}}{dI} = \frac{Q}{\sqrt{ - 2I}}\, .
\end{equation}
Furthermore one may define a variable $X$ by (\ref{NEM6}) and rewrite
Eq.~(\ref{NEM55d}),
\begin{equation}
\label{5d}
  -2 I \frac{d \mathcal{L}}{dI} = \frac{X}{r^3} \, .
\end{equation}
Because
\begin{equation}
\rho = - 2 I \frac{d \mathcal{L}}{dI} - \mathcal{L}\, ,
\end{equation}
instead of (\ref{NEM8}), we obtain
\begin{equation}
\label{NEM85d}
\mathcal{L} = \frac{X}{r^3} - \rho \, .
\end{equation}
Furthermore due to
\begin{equation}
\label{XI}
I\frac{d}{dI}= \frac{X}{2} \frac{d}{dX}\, ,
\end{equation}
Eq.~(\ref{5d}) gives
\begin{equation}
\label{5d2}
  - \frac{d \mathcal{L}}{dX} = \frac{1}{r^3} \, ,
\end{equation}
and therefore Eq.~(\ref{NEM8}) gives
\begin{equation}
\label{5d3}
\frac{d\rho}{dr} = \frac{1}{r^3}\frac{dX}{dr}
  - \frac{3X}{r^4} - \frac{dX}{dr}\frac{d \mathcal{L}}{dX} = - \frac{3X}{r^4}\, ,
\end{equation}
that is,
\begin{equation}
\label{NEM95d}
X = - \frac{r^4}{3} \frac{d\rho}{dr}\, .
\end{equation}
Therefore as in case of four dimensions, for the geometry expressed by
$\e^{2\nu(r)}$ in (\ref{GBiv}),
by using (\ref{GBvid2B}), we find the explicit $r$-dependence of $\rho$.
Then Eq.~(\ref{NEM95d}) makes to express $X$ in terms of $r$, which could be
solved with respect to
$r$ as $r=r \left( X \right)$.
Substituting the obtained expression $r=r \left( X \right)$ into
(\ref{NEM85d}), we obtain the
explicit form of $\mathcal{L}$, $\mathcal{L}=\mathcal{L} \left( X \right)$.
In case that one cannot solve Eq.~(\ref{NEM95d}) explicitly and/or there could
not be the
one-to-one correspondence between $X$ and $r$, we can use the Lagrangian
density corresponding to (\ref{LBCX}),
\begin{equation}
\label{LBCX5d}
\mathcal{L}_{BCX} \equiv \frac{X}{C^3} - \rho \left(r=C\right)
+ B \left( X + \frac{C^4}{3} \left. \frac{d\rho}{dr} \right|_{r=C} \right) \, .
\end{equation}

\subsection{An Example}

As in the case of four dimensions,
in the background of the Minkowski space-time or the anti-de Sitter space-time,
we may express $\e^{2\nu(r)}$ as in (\ref{NEM10}) and we require
$h(r)\to r_1 r_2 \cdots r_{2N} \left( 1 - \left( \sum_{i=1}^{2N} \frac{1}{r_i}
\right) r
+ \mathcal{O}\left( r^2 \right) \right)$ in the limit of $r\to 0$ and in the
limit of $r\to \infty$,
$h(r)$ behaves as $h(r)\to r^{2N}$ for the Minkowski background or
$h(r)\to \frac{r^{2N+2}}{l^2}$ with a length parameter $l$ for the anti-de
Sitter background.
In the de Sitter background, $h(r)$ is expressed as in (\ref{NEM11}) and we
require $h(r)\to - r_1 r_2 \cdots r_{2N-1}
\left( 1 - \left( \sum_{i=1}^{2N-1} \frac{1}{r_i} \right) r
+ \mathcal{O}\left( r^2 \right) \right)$ in the limit of $r\to 0$ and
$h(r)\to - \frac{r^{2N+1}}{l^2}$ with a length parameter $l$ in the limit of $r
\to \infty$.
The regular black hole in the anti-de Sitter space-time or the Minkowski
space-time has even-number of horizons.
On the other hand, the regular black hole in the de Sitter space-time has odd
number of horizons,
in which the largest horizon corresponds to the cosmological horizon.

As an example, we consider the regular black hole in (\ref{NEM12B}),
where there appear two horizons in the anti-de Sitter background.
Then by using (\ref{GBvid2B}) and (\ref{NEM95d}), we find
\begin{align}
\label{GBrho}
\rho =& \frac{3 \left\{ - 2 r^4 + \left( r_2 - r_1 \right)^2 r^2
+ 2 r_1 r_2 \left( r_1^2 + r_2^2 + r_1 r_2 \right) \right\}}
{2 r_1 r_2 \left( r^2 + r_1 r_2 \right)^2} \nn
& - \frac{12c \left\{ r^6 + \left( - r_1^2 - r_2^2 + r_1 r_2 \right) r^4
  -3 r_1r_2 \left( r_1^2 + r_2^2 + r_1 r_2 \right) r^2
+ r_1 r_2 \left( r_1^2 + r_2^2 + r_1 r_2 \right)^2\right\}}
{r_1^2 r_2^2 \left( r^2 + r_1 r_2 \right)^3} \, , \\
\label{GBX}
X =& - r^5 \left[ \frac{ \left( r_1 + r_2 \right)^2 \left( r^2 + 3 r_1 r_2
\right)}
{r_1 r_2 \left( r^2 + r_1 r_2 \right)^3}
+ \frac{8c \left( r_1 + r_2 \right)^2 \left\{ - r^4 - 4 r_1 r_2 r^2
+ 3 r_1 r_2 \left( r_1^2 + r_2^2 + r_1 r_2 \right) \right\}}
{r_1^2 r_2^2 \left( r^2 + r_1 r_2 \right)^4} \right] \, .
\end{align}
Then the Lagrangian density in (\ref{LBCX5d}) has the following form,
\begin{align}
\label{LBCXGB1}
\mathcal{L}_{BCX} =& \frac{X}{C^3}
  - \frac{3 \left\{ - 2 C^4 + \left( r_2 - r_1 \right)^2 C^2
+ 2 r_1 r_2 \left( r_1^2 + r_2^2 + r_1 r_2 \right) \right\}}
{2 r_1 r_2 \left( C^2 + r_1 r_2 \right)^2} \nn
& - \frac{12c \left\{ C^6 + \left( - r_1^2 - r_2^2 + r_1 r_2 \right) C^4
  -3 r_1r_2 \left( r_1^2 + r_2^2 + r_1 r_2 \right) C^2
+ r_1 r_2 \left( r_1^2 + r_2^2 + r_1 r_2 \right)^2\right\}}
{r_1^2 r_2^2 \left( C^2 + r_1 r_2 \right)^3} \nn
& + B \left[ X + C^5 \left\{ \frac{ \left( r_1 + r_2 \right)^2 \left( C^2 + 3
r_1 r_2 \right)}
{r_1 r_2 \left( C^2 + r_1 r_2 \right)^3}
+ \frac{8c \left( r_1 + r_2 \right)^2 \left\{ - C^4 - 4 r_1 r_2 C^2
+ 3 r_1 r_2 \left( r_1^2 + r_2^2 + r_1 r_2 \right) \right\}}
{r_1^2 r_2^2 \left( C^2 + r_1 r_2 \right)^4} \right\} \right] \, .
\end{align}
By using the redefinitions in (\ref{redefinitions}), we rewrite the Lagrangian
density (\ref{LBCXGB1}) as follows,
\begin{align}
\label{LBCXGB2}
\mathcal{L}_{BCX} =& \frac{X}{C^3}
  - \frac{3 \left\{ - 2 C^4 + \left( \alpha - 2 \right) C^2
+ 2 \left( \alpha + 1 \right) \right\}}
{2l^2 \left( C^2 + 1 \right)^2}
  - \frac{12c \left\{ C^6 + \left( - \alpha + 1 \right) C^4
  -3 \left( \alpha + 1 \right) C^2
+ \left( \alpha + 1 \right)^2\right\}}
{l^4 \left( C^2 + 1 \right)^3} \nn
& + B \left[ X + l C^5 \left\{ \frac{ \left( \alpha + 2 \right) \left( C^2 +
3 \right)}
{\left( C^2 + 1 \right)^3}
+ \frac{8c \left( \alpha + 2 \right) \left\{ - C^4 - 4 C^2
+ 3 \left( \alpha + 1 \right) \right\}}
{\left( C^2 + 1 \right)^4} \right\} \right] \, .
\end{align}
Then we find again that this model given by the Lagrangian density
(\ref{LBCXGB2}) has
only two coupling constants $\alpha$ and $l^2$.
In terms of $\alpha$ and $l^2$, the horizon radii $r_1$ and $r_2$ are given by
(\ref{r12}).
Therefore in order to consider the Nariai limit, where $r_1 \to r_2$, or,
$\alpha \to 2$,
we need to add the Lagrangian density in (\ref{Balpha}).
The Nariai limit is given by making $\alpha \to 2$ and redefining
(\ref{Nrlim2B}) and taking the limit of $\epsilon \to 0$.
By using the further redefinitions in (\ref{Nrlim5}),
we obtain the metric corresponding to (\ref{Nrlim6}),
\begin{equation}
\label{Nrlim6GB}
ds^2 = \frac{2}{l^2\cosh^2 \frac{\rho}{l^2}} \left(d\tau^2 - d\rho^2 \right)
+ l^2 \sum_{i,j=1}^3 \tilde g_{ij} dx^i dx^j\, .
\end{equation}

\subsection{Thermodynamics}

Let us investigate the thermodynamical properties of the obtained black hole solution
(\ref{NEM12B}).
In order to estimate the entropy, we use the Wald
formula \cite{Wald:1993nt,Iyer:1994ys,Iyer:1995kg}.
In general $D$ dimensional space-time, the formula is given by
\begin{equation}
\label{wald}
\mathcal{S}= - \frac{1}{8G} \int_\mathrm{horizon} d^{D-2}x \sqrt{h}
\frac{\delta \mathcal{L}_\mathrm{gravity}}{\delta R_{\mu\nu\rho\sigma} }
\epsilon^{\mu\nu} \epsilon_{\rho\sigma} \, .
\end{equation}
Here $G$ is the Newton constant, which we now define as $8\pi G = 1$.
$\epsilon_{\mu\nu}$ is the anti-symmetric tensor in the two dimensional
space-time
perpendicular to the horizon, that is, the space-time is given by $\mu,\nu=t,r$
and we choose $\epsilon_{tr}=1$.
Furthermore $ \mathcal{L}_\mathrm{gravity}$ is the Lagrangian density of the
gravity theory, which is now given by
\begin{align}
\label{Lgrav}
\mathcal{L}_\mathrm{gravity} = &
\frac{R}{2} + c \left( R^2 -4 R_{\mu\nu} R^{\mu\nu}
+ R_{\mu\nu\xi\sigma} R^{\mu\nu\xi\sigma} \right) \nn
= & \frac{1}{2} g^{\mu\rho} g^{\nu\sigma} R_{\mu\nu\rho\sigma}
+ c \left( g^{\mu\rho} g^{\nu\sigma} g^{\mu'\rho'} g^{\nu'\sigma'}
  - 4 g^{\mu\mu'}g^{\rho\rho'} g^{\nu\sigma} g^{\nu'\sigma'}
+ g^{\mu\mu'} g^{\nu\nu'} g^{\rho\rho'} g^{\sigma\sigma'} \right)
R_{\mu\nu\rho\sigma} R_{\mu'\nu'\rho'\sigma'} \, .
\end{align}
As the area of three-dimensional sphere is given by $2\pi^2 r^3$, we find,
\begin{equation}
\label{wald2}
\mathcal{S}= 2 \pi^3 r_2^3 \left.
\left\{ 1 + 2 c \left( 2 R - 4 \e^{-2\nu} R_{tt} + 4 \e^{2\nu} R_{rr}
+ 4 R_{trtr} \right) \right\}\right|_{r=r_2} \, .
\end{equation}
Then by using
\begin{align}
\label{Rtrtr}
R_{trtr} =&\e^{2\nu}\left(\nu'' + 2 {\nu'}^2 \right)
= \frac{r^6 + 3 r_1 r_2 r^2 + r_1 r_2 \left( 3 r_1^2 +3 r_2^2 + 9 r_1 r_2
\right) r^2
  - r_1^2 r_2^2 \left( r_1^2 + r_2^2 + r_1 r_2 \right)}{r_1 r_2 \left( r^2 + r_1
r_2 \right)^3}
\, ,\nn
R_{tt}=& \e^{4\nu} \left( \nu'' + 2 {\nu'}^2 + \frac{3\nu'}{r} \right)
= 4 \e^{2\nu} \left( \frac{ r^6 + 3 r_1 r_2 r^4 + 3 r_1^2 r_2^2 r^2
  - r_1^2 r_2^2 \left( r_1^2 + r_2^2 + r_1 r_2 \right)}{r_1 r_2 \left( r^2 + r_1
r_2 \right)^3}
\right) \, ,\nn
R_{rr}=& - \left( \nu'' + 2{\nu'}^2 + \frac{3\nu' }{r}\right)
= - 4 \e^{-2\nu} \left( \frac{ r^6 + 3 r_1 r_2 r^4 + 3 r_1^2 r_2^2 r^2
  - r_1^2 r_2^2 \left( r_1^2 + r_2^2 + r_1 r_2 \right)}{r_1 r_2 \left( r^2 + r_1
r_2 \right)^3}
\right) \, ,\nn
R=& \e^{2\nu}\left( - 2\nu'' - 4 {\nu'}^2 - \frac{12\nu'}{r}
+ \frac{6\e^{-2\nu} - 6}{r^2} \right) \nn
=& \frac{2 \left\{ - 10 r^6 + 3 \left( r_1^2 + r_2^2 - 8 r_1 r_2 \right) r^4
+ 3 r_1 r_2 \left( 3 r_1^2 + 3 r_2^2 - 4 r_1 r_2 \right) r^2
+ 10 r_1^2 r_2^2 \left( r_1^2 + r_2^2 + r_1 r_2 \right) \right\}}
{r_1 r_2 \left( r^2 + r_1 r_2 \right)^3} \, ,
\end{align}
one gets
\begin{equation}
\label{wald3}
\mathcal{S}= 2 \pi^3 r_2^3 \left[ 1 + \frac{ 8 c \left( 17 r_1^2 - 5 r_1 r_2 -
14 r_2^2 \right)}{r_1 r_2^2 \left( r_1 + r_2 \right)} \right]\, .
\end{equation}
With the help of (\ref{r1}) and (\ref{Tr2}),
\begin{equation}
\label{ent3GB}
\mathcal{S} = 2 \pi^3 \left(
2 \pi^2 l^4 T^2 + l^2 + 2 \pi l^2 T \sqrt{ \pi^2 l^4 T^2 + l^2} \right)
\left\{ 1 - \frac{8c \left( l^2 - 3 \pi^2 l^4 T^2 + 31 \pi l^2 T \sqrt{ \pi^2
l^4 T^2 + l^2}
\right)}{ l^2 \left( \pi l^2 T + \sqrt{ \pi^2 l^4 T^2 + l^2}\right) \sqrt{
\pi^2 l^4 T^2 + l^2} }
\right\} \, .
\end{equation}
Then if $T^2 < \left( 3 \pi l^2 \right)^{-1}$ and $c$ is positive and large
enough,
or even if $T^2 > \left( 3 \pi l^2 \right)^{-1}$, if $c$ is negative and large
enough, the entropy becomes negative.
The negative entropy in the 5-dimensional Gauss-Bonnet gravity was first
observed
in Ref.~\cite{Cvetic:2001bk} (for the study of black hole thermodynamics
in this case, see \cite{Clunan:2004tb,Nojiri:2002qn,Cho:2002hq,Nojiri:2001aj,
Cai:2001dz,Wiltshire:1985us}.
When the negative entropy appears, the corresponding black hole solution
becomes
unstable or there is an ambiguity in the definition of the entropy in general
\cite{Clunan:2004tb,Nojiri:2002qn} related with possible AdS/dS
transition \cite{Nojiri:2002qn}.
Because the internal energy $E$ is given by $dE=Td\mathcal{S}$ and
the free energy is also defined by $F=E-T\mathcal{S}$, we find
\begin{equation}
\label{FS}
\frac{dF}{dT}= - \mathcal{S} \, ,
\end{equation}
and therefore
\begin{align}
\label{FF}
F =& - \int dT \mathcal{S} \nn
=& - 2\pi^3\left[ l^2 T + \frac{2}{3} \pi^2 l^4 T^3 + \frac{2}{3}\left(
\frac{1}{\pi}
+ \pi l^2 T^2 \right)\sqrt{ \pi^2 l^4 T^2 + l^2} \right. \nn
& \left. - 8c \left\{T
+ \frac{28}{3}\pi^2 l^2 T^3 + \frac{2}{3}\left( \frac{20}{\pi l^2}
+ 14 \pi T^2 \right) \sqrt{ \pi^2 l^4 T^2 + l^2} \right\} \right] \, .
\end{align}
  From the viewpoint of AdS/CFT correspondence, the free energy should
correspond to that of the field theory on the boundary of the anti-de Sitter
space-time.
In the standard AdS/CFT correspondence
\cite{Maldacena:1997re,Witten:1998qj},
the conformal field theory (CFT) is super Yang-Mills (SYM) theory.
In the conformal field theory, all the particles are massless and also for the
SYM theory, we are usually assuming that all the particles are massless and
therefore the free energy $F$ should be proportional to the 4th power of
the temperature $T$,
that is $F\propto T^4$.
For example, in case of $\mathcal{N}=4$ $U(N)$ SYM model, the free energy is
given by \cite{Gubser:1998nz}
\begin{equation}
\label{FN$SYM}
F= - \left( \frac{3}{4} + \mathcal{O}\left( N^{-\frac{3}{2}}\right) \right)
\frac{\pi^2}{6} N^2 V_0 T^4\, .
\end{equation}
On the other hand, the free energy in (\ref{FF}) is not proportional to $T^4$.
In the high energy limit, the usual particles behave as massless
particles but for the large $T$, the free energy $F$ in (\ref{FF}) behaves as
\begin{equation}
\label{FF2}
F \sim - 2 \pi^3 \left( \frac{4}{3}\pi^2 l^4 - \frac{448}{3}c\pi^2 l^2 \right)
T^3\, ,
\end{equation}
that is, $F$ is proportional to the 3d power of $T$, which may suggest the
effectively 3 dimensional
field theory or non-relativistic theory on the AdS boundary.

Therefore one may conjecture the correspondence of our 5d non-singular
black hole with some system of the condensed matter, for example.
For large $T$, the entropy in (\ref{ent3GB}) behaves as
\begin{equation}
\label{ent3GB2}
\mathcal{S} \sim 8 \pi^5 l^4
\left( 1 - \frac{224c}{ l^2} \right) T^2\, ,
\end{equation}
and therefore the thermodynamical energy $E=F+TS$ behaves as
\begin{equation}
\label{EnergyGB}
E \sim \frac{16\pi^5 l^4}{3}\left( 1 - \frac{280 c}{l^2} \right) T^3\, .
\end{equation}
Therefore if $1 - \frac{280 c}{l^2}>0$, the specific heat $C=\frac{dE}{dT}$
becomes positive again, which is different from the case of the standard black hole.
This requests further deep investigation of non-singular black holes
thermodynamics in non-linear theories.

\section{Discussion \label{Sec6}}

In summary, we investigated the regular black holes with multi-horizons
in modified gravity with non-linear electromagnetism.
We presented several explicit examples of the actions which give the
solutions describing
the non-singular black hole space-time with multi-horizons in the
Einstein gravity, the $F(R)$ gravity, and the 5 dimensional Gauss-Bonnet
gravity when coupling with non-linear electromagnetism is present.
We also studied the thermodynamics of the obtained non-singular black
hole solutions and found the
explicit expressions for the temperature, the entropy, the thermodynamical
energy, and the free energy.
Although the temperature vanishes in the extremal limit where the radii of
the two horizons coincide as for the standard black hole with multi-horizons,
the larger
temperature corresponds to the larger horizon radius at 
least, for some examples.
This is different from the
standard black holes thermodynamics, where the larger temperature
corresponds to the smaller horizon radius.
In relation with the above observation, we also found that the
specific heat  often becomes positive for the large 
temperature,
which is also different from
the standard black holes, where the specific heat is negative.
We should also note that the thermodynamical energy is not identical with the
ADM mass.
Furthermore in the case of the Gauss-Bonnet gravity, the entropy can
become negative what may indicate to the instability of the corresponding
parameters region.

Note that there is big interest to the study of black holes thermodynamics
in $F(R)$ gravity and five-dimensional Einstein-Gauss-Bonnet gravity
coupled with (non)linear electromagnetism (see Refs.~\cite{Iyer:1995kg,Deng:2017abh,Zeng:2016aly,
Hendi:2015xya,DiazAlonso:2009ak,Camanho:2015zqa,Hendi:2014bba,Wei:2014hba,Xu:2013zpa,
Cai:2013qga,delaCruzDombriz:2012xy,Hendi:2010dz,Miskovic:2010ey,Anninos:2008sj,Brihaye:2008xu,
Dehghani:2006cu, Clunan:2004tb,Chakraborty:2015taq,Hendi:2010zza}).
It is clear that these works may be generalized and more complicated
non-singular black holes may be obtained in this case too.
One can also investigate the anti-evaporation phenomena for such black 
holes.
This will be done elsewhere.

Also, there exists some interest in the study of relation between the 
regular black holes and
the energy conditions \cite{Balart:2014jia,Neves:2014aba}.
As one can construct general type of regular black holes by using the 
formulation
in \cite{Chinaglia:2017uqd}, it could be interesting to clarify this 
relation in most general case.

\section*{Acknowledgements}

This work is supported (in part) by
MEXT KAKENHI Grant-in-Aid for Scientific Research on Innovative Areas
``Cosmic Acceleration'' (No. 15H05890) (SN) and by MINECO (Spain), project
FIS2013-44881, FIS2016-76363-P(SDO) and by CSIC I-LINK1019 Project
(SDO and SN).

\end{document}